\documentclass[aps,preprint,amsmath,amssymb,showpacs]{revtex4}

\usepackage{graphicx}
\begin{document}

\title{Polarized Single Top Quark Production in $e\gamma$ Collision and
Anomalous $Wtb$ Couplings}

\author{B. \c{S}ahin}
\email[]{dilec@science.ankara.edu.tr}
\affiliation{Department of Physics, Faculty of Sciences,
Ankara University, 06100 Tandogan, Ankara, Turkey}

\author{\.{I}. \c{S}ahin}
\email[]{isahin@science.ankara.edu.tr} \affiliation{Department of
Physics, Faculty of Sciences, Ankara University, 06100 Tandogan,
Ankara, Turkey}

\begin{abstract}
We investigate the potential of $e\gamma$ collisions to probe
anomalous $Wtb$ couplings via the polarized single top quark
production process $e^{+} \gamma \to t \bar{b} \bar{\nu_{e}}$. We
find 95$\%$ confidence level limits on the anomalous coupling
parameters $F_{2L}$ and $F_{2R}$ with an integrated luminosity of
$500 fb^{-1}$ and $\sqrt{s}=0.5, 1$ and $1.5$ TeV energies. The
effects of top quark spin polarization on the anomalous $Wtb$
couplings are discussed. It is shown that polarization leads to a
considerable improvement in the sensitivity limits.
\end{abstract}


\maketitle

\section{Introduction}

The standard model (SM) has been tested with good accuracy and it
has been proved to be successful in the energy scale of the present
colliders. However, it is generally believed that SM is embedded in
a more fundamental theory (new physics) in which its effects can be
observed at higher energy scales. The top quark is the heaviest
fermion in the SM. Its mass is at the electroweak symmetry-breaking
scale. Because of its large mass, the top quark and its couplings
are expected to be more sensitive to new physics than other
particles \cite{zhang}. Therefore precision measurements of top
quark couplings will be the crucial test of the structure of the SM.
A deviation of the couplings from the expected values would indicate
the existence of new physics beyond the SM.

In this work we analyzed anomalous $Wtb$ and $\gamma Wtb$ couplings
in the single top production process $e^{+} \gamma \to t \bar{b}
\bar{\nu_{e}}$. Since the top quark is very heavy, its weak decay
time is much shorter than the typical time for the strong
interactions to affect its spin \cite{bigi}. Therefore the
information on its polarization is not disturbed by hadronization
effects but transferred to the decay products. The angular
distribution of the top quark decay involves correlations between
top decay products and top quark spin:

\begin{eqnarray}
 \frac{1}{\Gamma_{T}}\frac{d\Gamma}{dcos\theta}=\frac{1}{2}(1+A_{\uparrow\downarrow}\alpha cos\theta)
\end{eqnarray}

Here the dominant decay chain of the top quark in the standard model
$t \to W^{+}b(W^{+} \to l^{+}\nu,\bar{d}u)$ is considered.
$A_{\uparrow\downarrow}$ is the spin asymmetry and $\theta$ is
defined as the angle between top quark decay products and the top
quark spin quantization axis in the rest frame of the top quark.
$\alpha$ is the correlation coefficient and $\alpha=1$ for $l$ or
$\bar{d}$, which leads to the strongest correlation. We take into
account top quark spin polarization along the direction of various
spin bases to improve the sensitivity limits.

Anomalous $Wtb$ and $\gamma Wtb$ couplings can be analyzed in a
model independent way by means of the effective Lagrangian approach
\cite{bmuller, renard, yang, ladinsky}. We consider the following
couplings, which are necessary for the process $e^{+} \gamma \to t
\bar{b} \bar{\nu_{e}}$.

\begin{eqnarray}
L&&=\frac{g_{w}}{\sqrt{2}} [W_{\mu}\bar{t}(\gamma^{\mu}F_{1L}P_{-}
+\gamma^{\mu}F_{1R}P_{+})b -\frac{1}{2m_{w}} W_{\mu\nu}
\bar{t}\sigma^{\mu\nu} (F_{2L}P_{-}+F_{2R}P_{+})b] + h.c.
\end{eqnarray}
where

\begin{eqnarray}
W_{\mu\nu}=D_{\mu}W_{\nu}-D_{\nu}W_{\mu} \;\;,\;\;
D_{\mu}=\partial_{\mu}-ieA_{\mu} \nonumber \\
P_{\mp}=\frac{1}{2}(1\mp\gamma_{5}),  \;\;,\;\;
\sigma^{\mu\nu}={i\over 2} (\gamma^{\mu}\gamma^{\nu}
-\gamma^{\nu}\gamma^{\mu})
\end{eqnarray}

In the SM, the (V-A) coupling $F_{1L}$ corresponds to the
Cabibbo-Kobayashi-Maskawa (CKM) matrix element $V_{tb}$, which is
very close to unity and $F_{1R}$, $F_{2L}$ and $F_{2R}$ are equal to
zero. The (V+A) coupling $F_{1R}$ is severely bounded by the CLEO
$b\rightarrow s\gamma$ data \cite{cleo} at a level such that it will
be out of reach at expected future colliders. Therefore we set
$F_{1L}=0.999$ and $F_{1R}=0$ as required by present data
\cite{pdg}. The magnetic type anomalous couplings are related to the
coefficients $C_{tW\Phi}$ and $C_{bW\Phi}$ \cite{yang} in the
general effective lagrangian by

\begin{eqnarray}
F_{2L}={{C_{tW\Phi}\sqrt{2}v m_{w} }\over{\Lambda^{2}g}} \;\;\;\;
F_{2R}={{C_{bW\Phi}\sqrt{2}v m_{w} }\over{\Lambda^{2}g}}
\end{eqnarray}
where $\Lambda$ is the scale of new physics. Natural values of the
couplings $F_{2L(R)}$  are in the region \cite{zhang}  of
\begin{eqnarray}
{\sqrt{m_{b}m_{t}}\over {v}}\sim 0.1
\end{eqnarray}
and do not exceed unitarity violation bounds for $|F_{2L(R)}|\sim
0.6$ \cite{renard}.

There are many detailed discussions in the literature for $Wtb$
couplings in the single and pair top quark production. The single
top quark production cross section for the process $e^{+}e^{-} \to
Wtb$ has been discussed below and the above the $t\bar{t}$ threshold
\cite{mele} and for the process $e^{+}e^{-} \to e\bar{\nu}tb$ at
CERN LEP2 \cite{tanaka} and linear $e^{+}e^{-}$ collider
\cite{kolodziej} energies. Pair top production processes for a
future linear collider have been investigated in $e^{+}e^{-}$ and
$\gamma\gamma$ collisions \cite{grzadkowski}. $Wtb$ couplings have
also been investigated at Fermilab Tevatron and CERN LHC
\cite{dicus, ohl}. In $ep$ collision, the $Wtb$ couplings were
analyzed for polarized top quarks via the process $ep \to
t\bar{b}\bar{\nu}+X$ \cite{atag}. It was shown that polarization
leads to a significant improvement in the sensitivity limits. In the
literature there have been several studies of anomalous $Wtb$
couplings in $e\gamma$ collisions \cite{boos}. Different from these
studies we take into account top quark spin polarization along the
direction of various spin bases to improve the sensitivity limits.

\section{Cross Sections of Polarized Top Quarks in the $e\gamma$ Collision}

Research and development of linear $e^{+}e^{-}$ colliders have been
progressing and the physics potential of these future machines is
under study. After linear colliders have been constructed their
operating modes of $e\gamma$ and $\gamma\gamma$ are expected to be
designed \cite{akerlof,barklow}. A real gamma beam is obtained
through Compton backscattering of laser light off a linear electron
beam, where most of the photons are produced at the high energy
region. The luminosities for $e\gamma$ and $\gamma\gamma$ collisions
turn out to be of the same order as the one for $e^{+}e^{-}$
\cite{Ginzburg}, so the cross sections for photoproduction processes
with real photons are considerably larger than the virtual photon
case. In our calculations we consider three different center of mass
energies $\sqrt{s}$=0.5, 1 and 1.5 TeV of the parental linear
$e^{+}e^{-}$ collider.

The spectrum of the backscattered photons is given by
\cite{Ginzburg}. We have

\begin{eqnarray}
f_{\gamma/e}(y)={{1}\over{g(\zeta)}}[1-y+{{1}\over{1-y}}
-{{4y}\over{\zeta(1-y)}}+{{4y^{2}}\over {\zeta^{2}(1-y)^{2}}}]
\end{eqnarray}

where

\begin{eqnarray}
g(\zeta)=&&(1-{{4}\over{\zeta}}
-{{8}\over{\zeta^{2}}})\ln{(\zeta+1)}
+{{1}\over{2}}+{{8}\over{\zeta}}-{{1}\over{2(\zeta+1)^{2}}}
\end{eqnarray}

with $\zeta=4E_{e}E_{0}/M_{e}^{2}$. $E_{0}$ is the energy of the
initial laser photon and $E_{e}$ is the energy of the initial
electron beam before Compton backscattering. $y$ is the fraction
that represents the ratio of the scattered photon and initial
electron energy for the backscattered photons moving along the
initial electron direction. The maximum value of $y$ reaches 0.83
when $\zeta=4.8$, in which case the backscattered photon energy is
maximized without spoiling the luminosity. The integrated cross
section over the backscattered photon spectrum is given by

\begin{eqnarray}
\sigma(s)=\int_{y_{min}}^{0.83}
f_{\gamma/e}(y)\hat{\sigma}(\hat{s}) dy
\end{eqnarray}

where $y_{min}=\frac{m_{t}^{2}}{s}$ and $\hat{s}$ is the square of
the center of mass energy of the subprocess $e^{+} \gamma \to t
\bar{b} \bar{\nu_{e}}$. $\hat{s}$ is related to $s$, the square of
the center of mass energy of  $e^{+}e^{-}$, by $\hat{s}=ys$.

\begin{figure}
\includegraphics{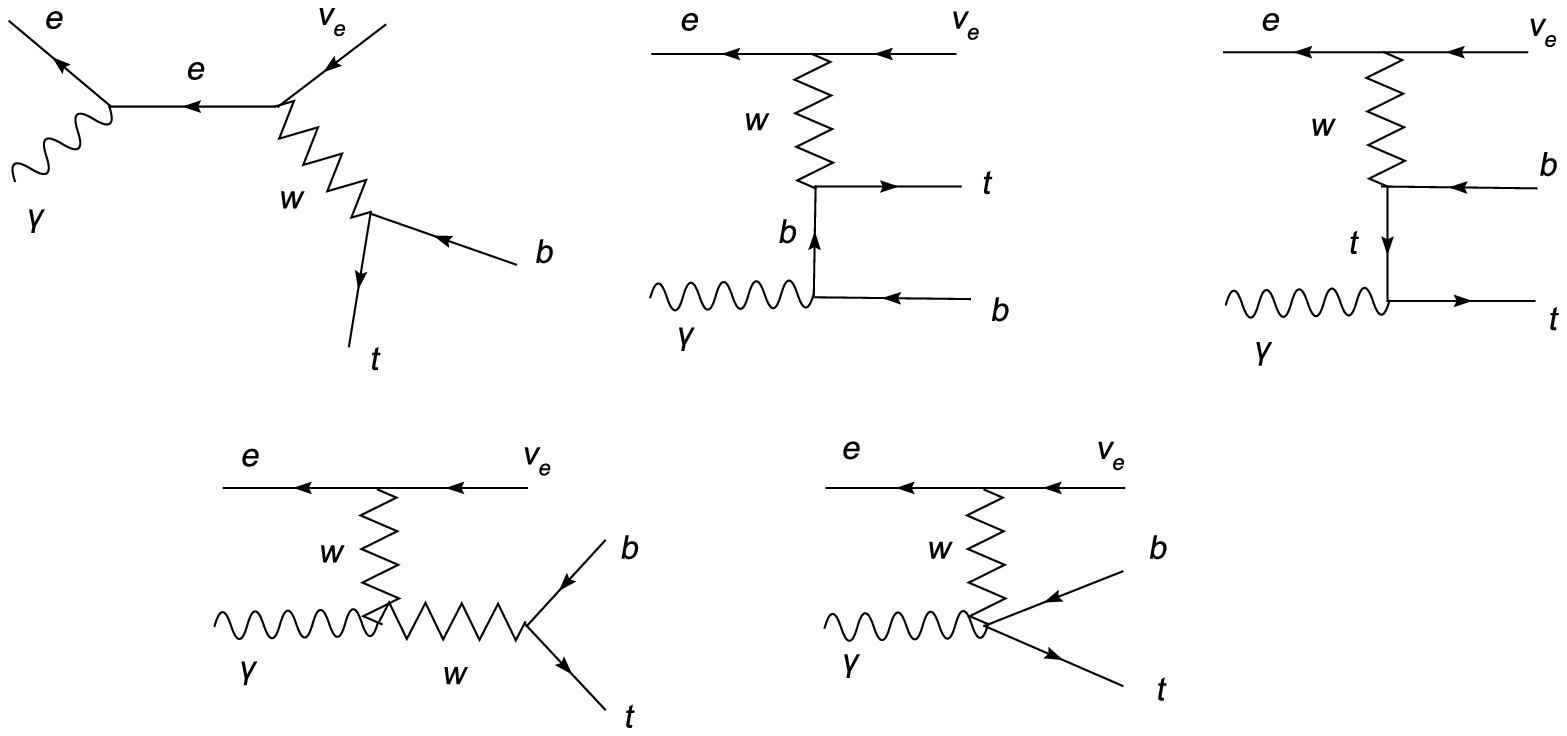}
\caption{Tree level Feynmann diagrams for the process $e^{+} \gamma
\to t \bar{b} \bar{\nu_{e}}$. \label{fig1}}
\end{figure}

In the SM  single production of the top quark via the process $e^{+}
\gamma \to t \bar{b} \bar{\nu_{e}}$ is described by four tree level
diagrams. Each of the diagrams contains a Wtb vertex and, due to its
V-A structure, the top quarks produced are highly polarized. It was
shown in ref.\cite{sahin} that the top quark possesses a high degree
of spin polarization when its spin decomposition axis is along the
incoming $e^{+}$ beam. In the effective Lagrangian approach, there
are five tree level diagrams; one of them contains an anomalous
$\gamma Wtb$ vertex, which is absent in the SM (Fig.\ref{fig1}).

The top quark possesses a large mass, so its helicity is frame
dependent and changes under a boost from one frame to another. The
helicity and chirality states do not coincide with each other and
there is no reason to believe that the helicity basis will give the
best description of the spin of top quarks. Therefore it is
reasonable to study other spin bases better than helicity for the
top quark spin.

The spin four-vector of a top quark is defined by

\begin{eqnarray}
s_{t}^{\mu}=(\frac{\vec{p}_{t}\cdot \vec{s^{\prime}}}{m_{t}} \,,\,
\vec{s^{\prime}}+\frac{\vec{p}_{t}\cdot \vec{s^{\prime}}
}{m_{t}(E_{t}+m_{t})}\vec{p}_{t})
\end{eqnarray}
where $(s_{t}^{\mu})_{RF}=(0,\vec{{s}^{\prime}})$ in the top quark
rest frame. Top quark spinors are the eigenstates of the operator
$\gamma_{5}(\gamma_{\mu}s_{t}^{\mu})$:

\begin{eqnarray}
\left[\gamma_{5}(\gamma_{\mu}s_{t}^{\mu})\right] \,u(p_{t},\pm
s)=\pm u(p_{t},\pm s)
\end{eqnarray}
Using eq.(10) one can easily obtain the spin projection operator:

\begin{eqnarray}
\hat{\Sigma}(s)=\frac{1}{2}(1+\gamma_{5}(\gamma_{\mu}s_{t}^{\mu}))
\end{eqnarray}
Therefore during amplitude calculations one should project the top
quark spin onto a given spin direction. We consider four different
top spin directions in the laboratory frame: the incoming positron
beam, the photon beam directions and the outgoing $\bar{b}$
direction and also the helicity basis.

The definition of the spin axis in the rest frame of the top quark
does not depend on the coordinate frame in which the cross section
is taken. So it is more convenient to calculate the cross section in
the $e^{+}e^{-}$ center of mass system (laboratory frame). In the
top quark rest frame, its spin direction along any beam (positron,
photon or $\bar{b}$ beam) can be defined as follows:

\begin{eqnarray}
  \vec{{s}^{\prime}}=\lambda \frac{\vec{{p}^{\star}}}
  {|\vec{{p}^{\star}}|} ,\,\,\,\,\, \lambda=\pm 1.
\end{eqnarray}
Here, $\vec{{p}^{\star}}$ is the particle momentum (positron, photon
or $\bar{b}$), observed in the rest frame of the top quark. Since
the particle momentum $\vec{p}$  is first defined in the
$e^{+}e^{-}$ center of mass system in which the cross section is
calculated, $\vec{{p}^{\star}}$ should be obtained by a Lorentz
boost from the $e^{+}e^{-}$ cm system:

\begin{eqnarray}
\vec{{p}^{\star}}=\vec{p}+\frac{\gamma-1}{\beta^{2}}
(\vec{\beta}\cdot \vec{p})\vec{\beta}
 -E\gamma \vec{\beta}
\end{eqnarray}
where $\vec{\beta}$ is the velocity of the top quark in the
$e^{+}e^{-}$ cm system. In the cross section calculations we have
performed a boost to obtain  $\vec{{p}^{\star}}$ at each point in
phase space.

\begin{figure}
\includegraphics{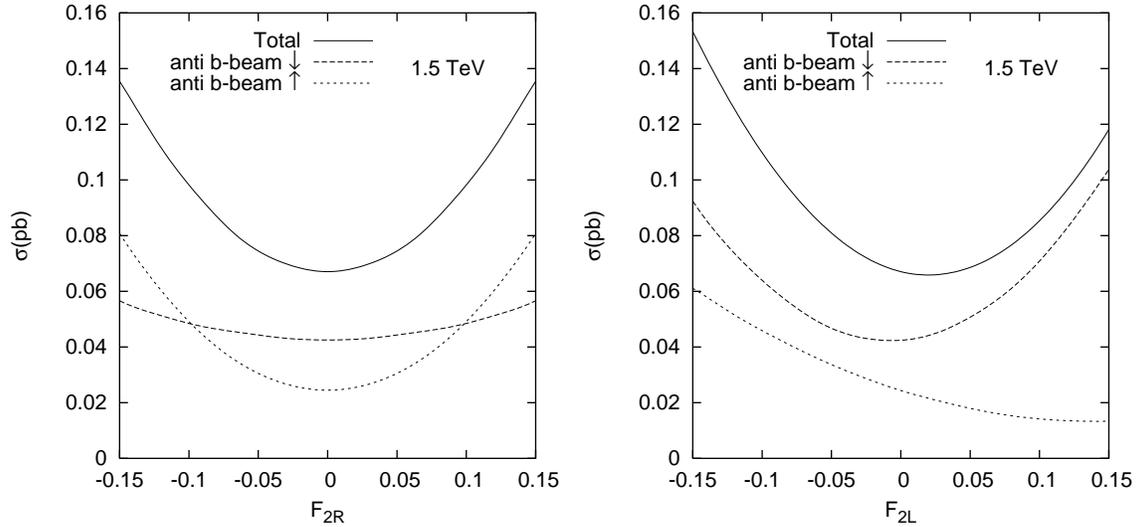}
\caption{The integrated cross section of the process $e^{+} \gamma
\to t \bar{b} \bar{\nu_{e}}$ as a function of the anomalous
couplings $F_{2R}$ and $F_{2L}$ at center of mass energy
$\sqrt{s}=1.5$ TeV of the parental linear $e^{+}e^{-}$ collider. The
top quark spin decomposition axis is along the $\bar{b}$-beam.
\label{fig2}}
\end{figure}

\begin{figure}
\includegraphics{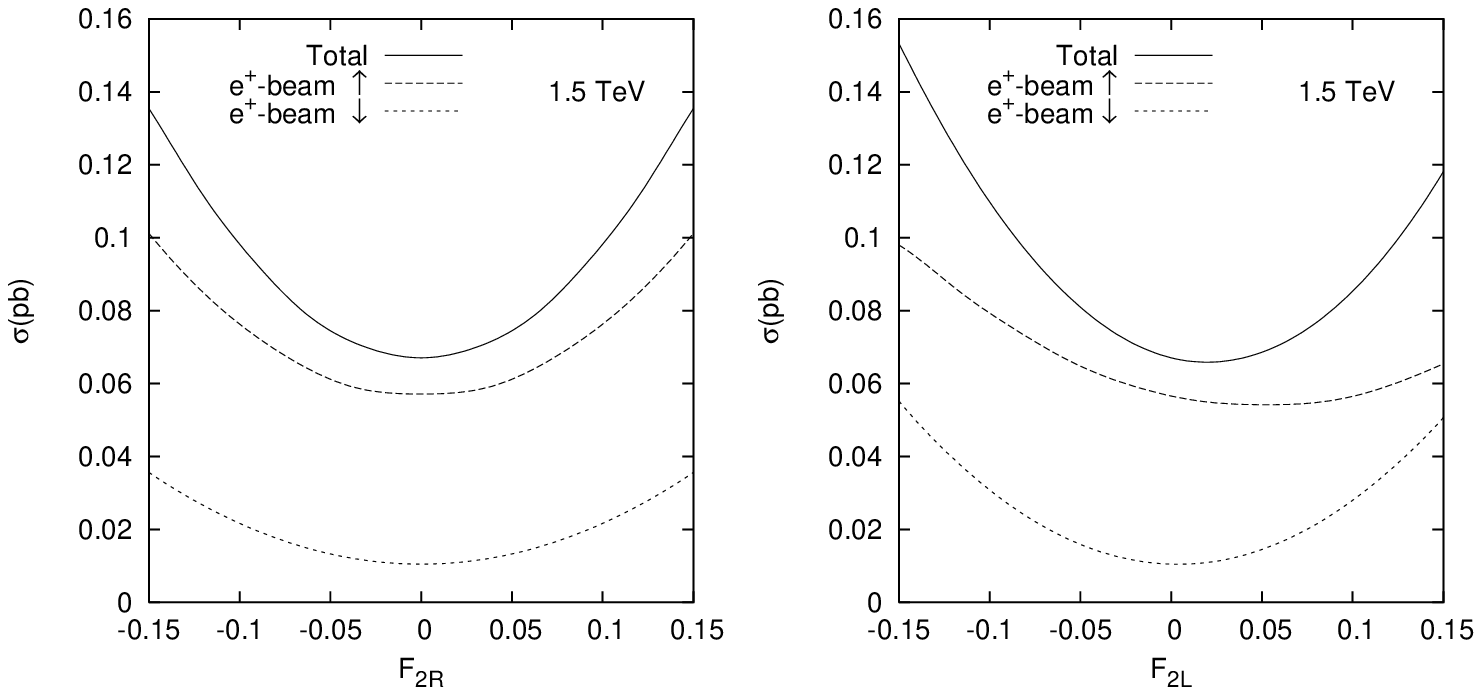}
\caption{The same as Fig. 2, but the top quark spin decomposition
axis is along the $e^{+}$-beam. \label{fig3}}
\end{figure}

One can see from Fig.\ref{fig2}-\ref{fig5} the influence of the top
quark spin polarizations on the deviations of the total cross
sections from their SM value at $\sqrt{s}=1.5$ TeV. In these figures
an up arrow $\uparrow$ (down arrow $\downarrow$) stands for spin up
$\lambda=+1$ (spin down $\lambda=-1$), and "L" and "R" represent
left and right helicity. These figures show that cross sections have
a symmetric behavior as a function of the anomalous parameter
$F_{2R}$. We see from Fig.\ref{fig4} that a polarized cross section
is almost insensitive to the anomalous parameter $F_{2R}$ at the
$\gamma$-beam $\downarrow$ spin polarization configuration. On the
other hand the cross section at this polarization configuration is
very sensitive to the anomalous parameter $F_{2L}$. Therefore the
$\gamma$-beam $\downarrow$ spin polarization configuration can be
used to isolate the anomalous coupling parameter $F_{2L}$.

In our calculations phase space integrations have been performed by
GRACE \cite{grace}, which uses a Monte Carlo routine.

\section{Angular Correlations Between Top Decay Products And Top Quark Spin}
Angular distributions of the top quark decay products have
correlations with its spin polarizations. Let us consider the
differential cross section for the complete process including
subsequent top decay,

\begin{eqnarray}
d\sigma \left(e^{+}\gamma \to t\bar{b} \bar{\nu_{e}} \to b \ell^{+}
\nu_{\ell} \bar{b} \bar{\nu_{e}}
\right)=&&\frac{1}{2s}|M|^{2}\frac{d^{3}p_{3}}{(2\pi)^{3}2E_{3}}\frac{d^{3}p_{4}}{(2\pi)^{3}2E_{4}}
\frac{d^{3}p_{5}}{(2\pi)^{3}2E_{5}}\frac{d^{3}p_{6}}{(2\pi)^{3}2E_{6}}\frac{d^{3}p_{7}}{(2\pi)^{3}2E_{7}}
\nonumber \\
&&\times(2\pi)^{4}\delta^{4}\left(\sum_{i}p_{i}-\sum_{f}p_{f}
\right)
\end{eqnarray}
where $p_{i}=p_{1},p_{2}$ are the momenta of the incoming fermions
and $p_{f}=p_{3},p_{4},p_{5},p_{6},p_{7}$ are the momenta of the
outgoing fermions. $|M|^{2}$ is the square of the full amplitude,
which is averaged over the initial spins and summed over the final
spins. The full amplitude can be expressed as follows:

\begin{eqnarray}
|M|^{2}(2\pi)^{4}\delta^{4}\left(\sum_{i}p_{i}-\sum_{f}p_{f}\right)=&&\int\frac{d^{4}q}{(2\pi)^{4}}
\left|\sum_{s_{t}}M_{a}(s_{t})D_{t}(q^{2})M_{b}(s_{t})\right|^{2}\nonumber
\\
&&\times(2\pi)^{4}\delta^{4}\left(p_{1}+p_{2}-p_{3}-p_{4}-q\right)\nonumber
\\ &&\times(2\pi)^{4}\delta^{4}\left(q-p_{5}-p_{6}-p_{7}\right)
\end{eqnarray}
Here q and $s_{t}$ are the internal momentum and spin of the top
quark. $D_{t}(q^{2})$ is the Breit-Wigner propagator factor. It is
given by

\begin{eqnarray}
D_{t}(q^{2})=\frac{1}{q^{2}-m_{t}^{2}+im_{t}\Gamma_{t}}
\end{eqnarray}
$M_{a}(s_{t})$ is the amplitude for the process $e^{+} \gamma \to t
\bar{b} \bar{\nu_{e}}$ with an on shell t quark. $M_{b}(s_{t})$ is
the decay amplitude for $t \to b \ell^{+} \nu_{\ell}$. The square of
the decay amplitude summed over the final fermion spins is given by

\begin{eqnarray}
|M_{b}(s_{t})|^{2}=\frac{2g_{w}^{4}}{[(p_{t}-p_{b})^{2}-m_{w}^{2}]^{2}}(p_{b}\cdot
p_{t}-p_{b}\cdot p_{\ell})(p_{\ell}\cdot p_{t}-m_{t}(s_{t}\cdot
p_{\ell}))
\end{eqnarray}
By means of this amplitude one can easily obtain equation (1), the
angular distribution of top quark decay.

Therefore, the full cross section has been written as a product of
production and decay parts. One can show that interference terms
from different spin states will vanish after integrating the decay
part over azimuthal angles of top quark decay products. Then the
following result can be reached:

\begin{eqnarray}
d\sigma\left(e^{+}\gamma \to t\bar{b} \bar{\nu_{e}} \to b \ell^{+}
\nu_{\ell} \bar{b} \bar{\nu_{e}}
\right)=\left[d\sigma\left(e^{+}\gamma \to \uparrow t\bar{b}
\bar{\nu_{e}}\right)\frac{d\Gamma\left(\uparrow t \to b \ell^{+}
\nu_{\ell}\right)}{\Gamma\left(t \to b \ell^{+} \nu_{\ell}\right)}
\right. \nonumber \\ \left. +d\sigma\left(e^{+}\gamma \to \downarrow
t\bar{b} \bar{\nu_{e}}\right)\frac{d\Gamma\left(\downarrow t \to b
\ell^{+} \nu_{\ell}\right)}{\Gamma\left(t \to b \ell^{+}
\nu_{\ell}\right)}\right]BR\left(t \to b \ell^{+} \nu_{\ell}\right)
\end{eqnarray}
where $BR\left(t \to b \ell^{+} \nu_{\ell}\right)$ is the leptonic
branching ratio for the top quark. Up and down arrows indicate the
spin up and spin down cases along a specified spin quantization
axis, respectively. $d\Gamma \left(\uparrow t \to b \ell^{+}
\nu_{\ell}\right)$ and $d\Gamma \left(\downarrow t \to b \ell^{+}
\nu_{\ell}\right)$ are differential decay rates for polarized top
quarks. The unpolarized rate is given by; $d\Gamma \left(t \to b
\ell^{+} \nu_{\ell}\right)=d\Gamma \left(\uparrow t \to b \ell^{+}
\nu_{\ell}\right)+d\Gamma \left(\downarrow t \to b \ell^{+}
\nu_{\ell}\right)$.

Top quark polarization can be determined by measuring the angular
distribution of outgoing charged lepton in the top rest frame. It is
possible to obtain from the expression (18) the polarized production
cross section as a coefficient of the angular distribution by a
fitting procedure. In this paper we ignore the problems associated
with the reconstruction of the top rest frame. We assume that the
top quark rest frame can be reconstructed.

\section{Sensitivity To Anomalous Couplings}

\begin{figure}
\includegraphics{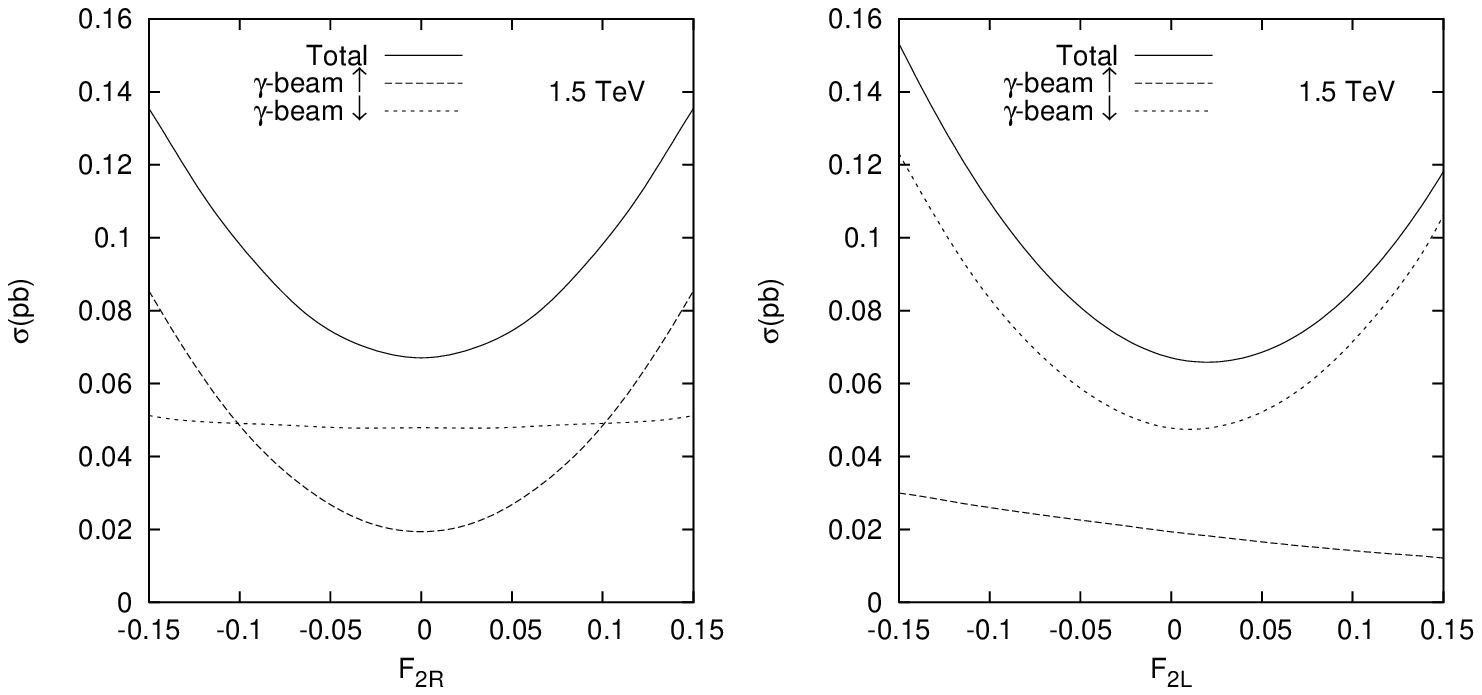}
\caption{The same as Fig. 3, but the top quark spin decomposition
axis is along the $\gamma$-beam. \label{fig4}}
\end{figure}

\begin{figure}
\includegraphics{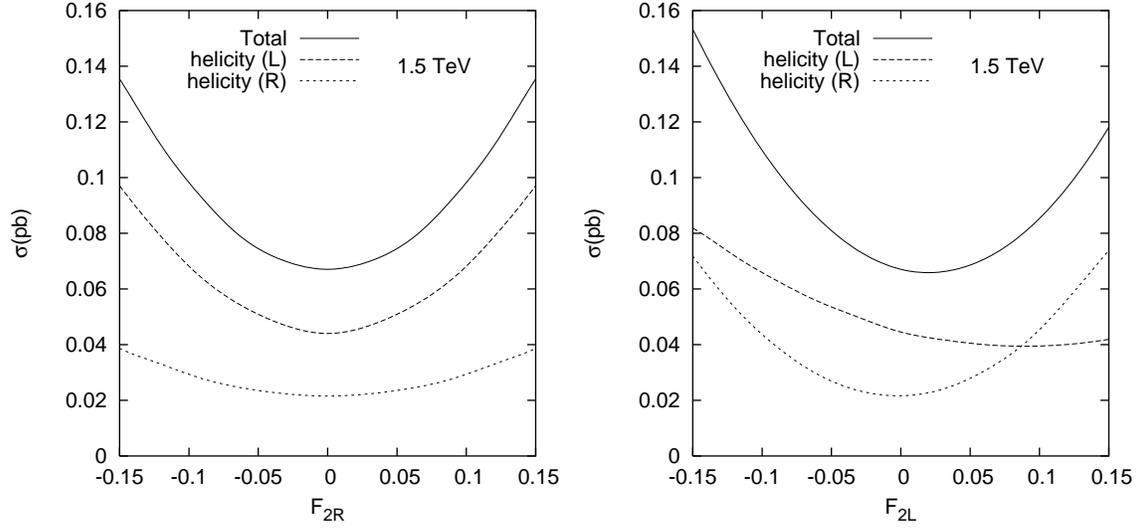}
\caption{The same as Fig. 4, but for the top quark helicity
basis.\label{fig5}}
\end{figure}

We have obtained 95\% C.L. limits on the anomalous coupling
parameters $F_{2L}$ and $F_{2R}$ using a $\chi^{2}$ analysis at
$\sqrt{s}=0.5, 1$ and $1.5$ TeV and an integrated luminosity
$L_{int}=500 fb^{-1}$ without systematic errors. The expected number
of events has been calculated considering the leptonic channel of
the W boson as the signal $N=AL_{int}\sigma B(W \to l\nu)$, where
$A$ is the overall acceptance. The limits for the anomalous coupling
parameters are given in Table \ref{tab1}-\ref{tab3} for top quark
spin polarization along the direction of various spin bases with the
acceptance $A=0.85$. One can see from these tables that the
sensitivity to the anomalous parameter $F_{2R}$ at the $\gamma$-beam
$\downarrow$ spin polarization configuration is the worst. This
feature is reflected in Fig.\ref{fig4} too. On the other hand, the
limits on $F_{2R}$ are most sensitive at the $\gamma$-beam
$\uparrow$ spin polarization configuration. The $\gamma$-beam
$\uparrow$ improves the sensitivity limits by a factor of 1.4 at
$\sqrt{s}=0.5$ TeV and by a factor of 1.5 at $\sqrt{s}=1$ TeV when
compared to the unpolarized (total) case.

Lower and upper bounds on the anomalous parameter $F_{2L}$ are not
symmetric, as can be seen from the tables. Polarization leads to a
significant improvement to these sensitivity bounds; the
$\bar{b}$-beam $\uparrow$ polarization configuration improves the
lower bounds on $F_{2L}$ by a factor of 2.7 at $\sqrt{s}=0.5$ TeV
and by a factor of 2 at $\sqrt{s}=1$ TeV when compared with the
unpolarized (total) case. At $\sqrt{s}=0.5$ TeV, $\bar{b}$-beam
$\downarrow$ and helicity right improves the upper bound on $F_{2L}$
by a factor of 1.75. These polarization configurations as well as
$\gamma$-beam $\uparrow$ leads to an improvement on the upper bound
by a factor of 3.5 at $\sqrt{s}=1$ TeV. The most sensitive bounds
are obtained at $\sqrt{s}=1.5$ TeV. $\gamma$-beam $\uparrow$ and
helicity right polarizations improve the upper bounds on $F_{2L}$ by
a factor of 5 when compared with the unpolarized case.

As a conclusion, we have obtained a considerable improvement in the
sensitivity bounds by taking into account top quark spin
polarization. Improved results by spin polarization in $e\gamma$
colliders with a luminosity of 500 $fb^{-1}$ have a higher potential
to probe the $F_{2L}$ and $F_{2R}$ couplings than Tevatron and CERN
LHC \cite{ohl} and also than the $ep$ collider TESLA+HERAp option
\cite{atag}. Furthermore, the linear $e^{+}e^{-}$ collider and its
$e\gamma$ mode provide a clean environment and the experimental
clearness is an additional advantage of $e\gamma$ collisions with
respect to $pp$, $p\bar{p}$ and $ep$ collisions.


\pagebreak


\begin{table}
\caption{ Sensitivity of the $e\gamma$ collision to anomalous
couplings at 95\% C.L. for the decomposition axis of the top quark
spin along the $e^{+}$-beam, $\gamma$-beam, $\bar{b}$-beam and
helicity directions. Only one of the couplings is assumed to deviate
from the SM at a time. $\sqrt{s}=0.5$ TeV. \label{tab1}}
\begin{ruledtabular}
\begin{tabular}{ccc}
Spin &$F_{2L}$ & $F_{2R}$\\
\hline
\hline
$e^{+}$-beam &   & \\
Up &  -0.10, 0.08  & -0.10, 0.10\\
Down & -0.06, 0.09   & -0.05, 0.05 \\
\hline
$\gamma$-beam &  & \\
Up &  -0.04, 0.05 & -0.05, 0.05\\
Down & -0.11, 0.05& -0.15, 0.15\\
\hline
$\bar{b}$-beam & & \\
Up & -0.03, 0.25  & -0.07, 0.07\\
Down & -0.22, 0.04& -0.1, 0.1 \\
\hline
Helicity &   & \\
Right & -0.16, 0.04 &-0.09, 0.09 \\
Left & -0.05, 0.21 &-0.08, 0.08 \\
\hline
Unpol& -0.08, 0.07   &-0.07, 0.07 \\
\end{tabular}
\end{ruledtabular}
\end{table}

\begin{table}
\caption{The same as table I, but for $\sqrt{s}=1$ TeV.
\label{tab2}}
\begin{ruledtabular}
\begin{tabular}{ccc}
Spin &$F_{2L}$ & $F_{2R}$\\
\hline \hline
$e^{+}$-beam &   & \\
Up &-0.02, 0.11 &-0.04, 0.04 \\
Down &-0.02, 0.04 &-0.03, 0.03  \\
\hline
$\gamma$-beam &  & \\
Up &-0.02, 0.02 &-0.02, 0.02 \\
Down &-0.02, 0.04 &-0.10, 0.10 \\
\hline
$\bar{b}$-beam & & \\
Up &-0.01, 0.30 &-0.03, 0.03 \\
Down &-0.05, 0.02 &-0.05, 0.05  \\
\hline
Helicity &   & \\
Right &-0.04, 0.02 &-0.04, 0.04 \\
Left &-0.01, 0.21 &-0.03, 0.03 \\
\hline
Unpol&-0.02, 0.07 &-0.03, 0.03 \\
\end{tabular}
\end{ruledtabular}
\end{table}

\begin{table}
\caption{The same as table II, but for $\sqrt{s}=1.5$ TeV.
\label{tab3}}
\begin{ruledtabular}
\begin{tabular}{ccc}
Spin &$F_{2L}$ & $F_{2R}$\\
\hline \hline
$e^{+}$-beam &   & \\
Up &-0.02, 0.12 &-0.03, 0.03 \\
Down &-0.01, 0.02 &-0.02, 0.02  \\
\hline
$\gamma$-beam &  & \\
Up &-0.02, 0.01 &-0.02, 0.02 \\
Down &-0.01, 0.03 &-0.11, 0.11 \\
\hline
$\bar{b}$-beam & & \\
Up &-0.01, 0.28 &-0.02, 0.02 \\
Down &-0.03, 0.02 &-0.04, 0.04  \\
\hline
Helicity &   & \\
Right &-0.02, 0.01 &-0.03, 0.03  \\
Left &-0.01, 0.18 &-0.02, 0.02 \\
\hline
Unpol&-0.01, 0.05 &-0.02, 0.02 \\
\end{tabular}
\end{ruledtabular}
\end{table}

\end{document}